\documentclass[10pt]{article}
\usepackage[margin=0.65in]{geometry}

\usepackage{verbatim}
\usepackage{amsmath}
\usepackage{breqn}
\usepackage{amssymb}
\usepackage{multirow}
\usepackage{import}
\usepackage[nomessages]{fp}

\usepackage{xcolor}
\usepackage{tabu}
\usepackage{float}
\usepackage{graphicx}

\usepackage[utf8]{inputenc}
\usepackage{gensymb}
\usepackage[section]{placeins}

\usepackage[style=nature]{biblatex}
\usepackage{lineno}
\usepackage{authblk}
\usepackage{textcomp}

\usepackage{titlesec}
\titleformat*{\section}{\normalsize\bfseries}
\titleformat*{\subsection}{\normalsize\bfseries}

\usepackage{setspace}
\usepackage{etoolbox}
\preto\longtable{\par\singlespacing}

\usepackage{hyperref}
\hypersetup{
	colorlinks = true,
	linkcolor = {red},
	citecolor = {blue},
	linkbordercolor = {white},
	citebordercolor = {purple}
}

\date{}

\title{Physical properties of asteroid Dimorphos as derived from the DART impact}

\author[1]{S.D. Raducan\footnote{sabina.raducan@unibe.ch}}%ORCID: 0000-0002-7478-0148
\author[1]{M. Jutzi} % 0000-0002-1800-2974
\author[2]{A.F. Cheng} 
\author[3]{Y. Zhang}
\author[2]{O. Barnouin} % 0000-0002-3578-7750
\author[4]{G. S. Collins} % 0000-0002-6087-6149
\author[2]{R. T. Daly} % 0000-0002-1320-2985
\author[4]{T. M. Davison}
\author[2]{C.M. Ernst} % 0000-0002-9434-7886
\author[3]{T. L. Farnham} 
\author[5]{F. Ferrari} % 0000-0001-7537-4996
\author[6,7]{M. Hirabayashi} % 0000-0002-1821-5689 
\author[8]{K. M. Kumamoto} % 0000-0002-0400-6333
\author[9,10]{P. Michel} % 0000-0002-0884-1993
\author[11]{N. Murdoch} 
\author[6,7]{R. Nakano} 
\author[12]{M. Pajola} 
\author[13]{A. Rossi}
\author[9,3]{H. F. Agrusa} 
\author[14]{B. W. Barbee} 
\author[8]{M. Bruck Syal} 
\author[2]{N. L. Chabot} 
\author[15]{E. Dotto}
\author[16]{E. G. Fahnestock} 
\author[15]{P. H. Hasselmann} 
\author[17]{I. Herreros} 
\author[18]{S. Ivanovski} 
\author[19]{J.-Y. Li}
\author[12]{A. Lucchetti}
\author[20]{R. Luther}
\author[17]{J. Orm\"{o}}  
\author[8]{M. Owen}  
\author[21]{P. Pravec}  
\author[2]{A. S. Rivkin} 
\author[11]{C. Q. Robin}  
\author[22]{P. S\'{a}nchez}
\author[12]{F. Tusberti}  
\author[20]{K. W\"{u}nnemann} 
\author[23]{A. Zinzi} 
\author[15]{E. Mazzotta Epifani}  
\author[24]{C. Manzoni}  
\author[24]{B. H. May}

\affil[1]{Space Research and Planetary Sciences, Physics Institute, University of Bern, Switzerland}
\affil[2]{Johns Hopkins University Applied Physics Laboratory, Laurel, MD, USA}
\affil[3]{Department of Aerospace Engineering, University of Maryland, College Park, MD, USA}
\affil[4]{Department of Earth Science and Engineering, Imperial College London, UK}
\affil[5]{Department of Aerospace Science and Technology, Politecnico di Milano, Italy}
\affil[6]{Guggenheim School of Aerospace Engineering, Georgia Institute of Technology, Atlanta, GA, USA}
\affil[7]{Department of Aerospace Engineering, Auburn University, Auburn, AL, USA}
\affil[8]{Lawrence Livermore National Laboratory, Livermore, CA, USA}
\affil[9]{Université Côte dAzur, Observatoire de la Côte dAzur, CNRS, Laboratoire Lagrange, Nice, France}
\affil[10]{University of Tokyo, Department of Systems Innovation, School of Engineering, Tokyo, Japan}
\affil[11]{Institut Sup\'{e}rieur de l'A\'{e}ronautique et de l'Espace (ISAE-SUPAERO), Universit\'{e} de Toulouse, Toulouse, France}
\affil[12]{INAF-OAPD Astronomical Observatory of Padova, Italy}
\affil[13]{IFAC-CNR, Sesto Fiorentino (FI), Italy}
\affil[14]{NASA/Goddard Space Flight Center, Greenbelt, MD, USA}
\affil[15]{INAF-Osservatorio Astronomico di Roma, Italy}
\affil[16]{Jet Propulsion Laboratory, California Institute of Technology, Pasadena, CA, USA}
\affil[17]{Centro de Astrobiolog\'{i}a (CAB), CSIC-INTA, Carretera de Ajalvir km4, 28850 Torrejón de Ardoz, Spain}
\affil[18]{INAF-Osservatorio Astronomico di Trieste, Trieste, Italy}
\affil[19]{Planetary Science Institute, Tucson, AZ, USA}
\affil[20]{Museum f\"{u}r Naturkunde, Leibniz Institute for Evolution and Biodiversity Science, Berlin, Germany}
\affil[21]{Astronomical Institute of the Czech Academy of Sciences, Czech Republic}
\affil[22]{Colorado Center for Astrodynamics Research, University of Colorado Boulder, Boulder, CO, USA}
\affil[23]{Agenzia Spaziale Italiana; ASI Space Science Data Center, Italy}
\affil[24]{London Stereoscopic Company, London, UK}

\begin{document}

\maketitle
%\linenumbers

\newpage

{\bfseries{On September 26, 2022, NASA's Double Asteroid Redirection Test (DART) mission successfully impacted Dimorphos, the natural satellite of the binary near-Earth asteroid (65803) Didymos. Numerical simulations of the impact provide a means to explore target surface material properties and structures, consistent with the observed momentum deflection efficiency, ejecta cone geometry, and ejected mass. Our simulation, which best matches observations, indicates that Dimorphos is weak, with a cohesive strength of less than a few pascals (Pa), similar to asteroids (162173) Ryugu and (101955) Bennu. We find that a bulk density of Dimorphos, $\rho_B$, lower than $\approx$ 2400 kg/m$^3$, and a low volume fraction of boulders ($\lesssim$ 40 vol\%) on the surface and in the shallow subsurface}, are consistent with measured data from the DART experiment. These findings suggest Dimorphos is a rubble pile that might have formed through rotational mass shedding and re-accumulation from Didymos. Our simulations indicate that the DART impact caused global deformation and resurfacing of Dimorphos. ESA's upcoming Hera mission may find a re-shaped asteroid, rather than a well-defined crater. }

\section*{Introduction}
DART was a planetary defense mission to demonstrate the feasibility of using a kinetic impactor to change the trajectory of an asteroid  \cite{Daly2023}. The impact was successful and highly effective, resulting in a reduction in Dimorphos' orbital period around Didymos, which was initially 11 hours and 55 minutes, by 33 $\pm$ 1 minutes \cite{Thomas2023}. 
The LICIACube Unit Key Explorer (LUKE) instrument onboard the cubesat \cite{Dotto2021} captured images of the system between 29 and 320 seconds after impact, revealing filamentary streams of ejecta and other complex patterns that expanded for several kilometres from the impact site \cite{Dotto2023}. Moreover, the dramatic brightening of the Didymos system by solar illumination of released impact ejecta was observed from ground- and space-based telescopes \cite{Thomas2023, Li2023, Graykowski2023} for many weeks after the impact. 

% DART impact scenario
The obtained 33-minute reduction in the binary orbital period \cite{Thomas2023} implies a momentum transfer to Dimorphos that exceeded the incident momentum of the DART spacecraft by a factor $\beta$, ranging from 2.2 to 4.9, depending on the mass of Dimorphos \cite{Cheng2023} -- which is not currently known but will be measured by the ESA Hera spacecraft in early 2027 \cite{Michel2022}. The $\beta$ parameter is defined as the ratio of the target momentum increment after the impact to the impactor momentum, in the direction of the net ejecta momentum, and is related to the additional thrust from the production of impact ejecta \cite{Cheng2023, Rivkin2021}. $\beta$ strongly depends on impact conditions (i.e., impact velocity and impact angle \cite{Daly2023}) and target material properties, such as strength, porosity, bulk density and target surface structure \cite{HH2012, Raducan2019, Stickle2022}. 

The spacecraft information, location and impact angle for the DART impact are well understood \cite{Daly2023}. However, the mass and surface properties of Dimorphos are still ambiguous. It was not possible to directly measure the mass and bulk density of Dimorphos with DART or LICIACube. Instead, these parameters were estimated from the total mass of the binary system, derived from Dimorphos' orbit \cite{Pravec2022}, and updated volume estimates of Didymos and Dimorphos provided by DART \cite{Daly2023}. The estimated bulk density of Dimorphos ranges from $\rho_B$ = 1500 to 3300 kg/m$^3$ \cite{Daly2023, Cheng2023}.

The surface material properties and sub-surface structure of Dimorphos were also not directly measured. However, these target parameters are vital for understanding the impact process and transforming the kinetic impactor method from a full-scale experiment by DART into a well-understood and reliable mitigation technique for planetary defence. Moreover, knowledge of the material properties of Dimorphos relates to the origin and evolution of the Didymos-Dimorphos system, as well as the overall characteristics of rubble-pile asteroids and binary asteroid systems. 

In this work, we simulate the DART impact numerically and compare the results with observations to infer the properties of Dimorphos. We performed numerical simulations of the DART impact using the shock physics code {\it Bern SPH} \cite{Jutzi2008, Jutzi2015} over a range of assumed sets of material properties and interior structures for Dimorphos. We represented the DART spacecraft as a low-density spherical projectile of equivalent mass (see Projectile section in Methods), and the impact velocity and angle were also kept fixed. We simulated the asteroid's response to the DART impact for up to one hour after the time of impact, using the numerical approach developed in \cite{Raducan2022, Jutzi2022} to model late-stage, low-speed deformation (see Modelling approach for the late stage evolution in Methods). Bern SPH's fast-integration scheme was validated against laboratory experiments \cite{ormo2022} and recently applied successfully to the Hayabusa2 SCI impact modelling \cite{Jutzi2022}. 
Due to the relatively short timescales modelled, the rotation of Dimorphos around Didymos and Didymos’s gravity were not accounted for.

We obtain realistic configurations of boulders for our rubble-pile targets from simulations of gravitational collapse of a cloud of spherical particles with a predefined size-frequency distribution (SFD) \cite{Daly2023, Pajola2023}. To closely replicate the topography described by \cite{Daly2023}, we then selectively removed particles positioned near the surface.

To explore a large possible range of boulder volume fractions (0 vol\% to 50 vol\%), we replaced some of these boulders with matrix material when we built our target asteroid. We only explicitly modelled boulders larger than 2.5 m in diameter, and the space between boulders was filled with matrix material. Boulders smaller than 2.5 m were removed from the SFD because they were too small to be resolved individually. Thus, components smaller than 2.5 m were considered part of the matrix, which was modelled as a granular material with low, but limited cohesion. This approach created an asteroid whose interior structure is similar to its surface and whose structure overall is consistent with a gravitational collapse. For the purpose of this study, other deep interior structures are not considered.

%%%%%%%%%%%%%%%%
The bulk porosity of Dimorphos results from macroporosity between individual rocks and boulders as well as microporosity within rocks. Analysis of reflectance spectra of Didymos indicates that the best meteorite analogues for boulders on Dimorphos are L/LL ordinary chondrites \cite{deLeon2006, Dunn2013, Ieva2022}, which have grain densities of $\approx$ 3200--3600 kg/m$^3$, and low microporosities of $\approx$ 8--10\% \cite{Flynn2018}. Using the method described by \cite{Grott2020} (See Dimorphos macroporosity calculations in Methods), we calculated macroporosities of 38 $\pm$ 3\% from the boulder SFD in the last complete image taken by DART \cite{Daly2023} and 34 $\pm$ 4\% from the global SFDs measured on Dimorphos \cite{Pajola2023}. The derived macroporosity is primarily determined by the boulder SFD and the sphericity/roundness of the boulders. It is important to note that the macroporosity estimation is largely independent of the assumed minimum boulder size. In our simulations, the initial micropososity within boulders was fixed at 10\% and the initial porosity of the matrix (macroporosity + microporosity) was varied between 35\% and 65\%. Both the porosity in the boulders and in the matrix were modelled using the $P-\alpha$ porosity compaction model \cite{Jutzi2008}. 

Based on laboratory measurements of meteorite falls, the average tensile strength of ordinary chondrites is 24 $\pm$ 11 MPa, with no statistical difference between L and LL types \cite{Flynn2018}. 
In all our simulations, the initial material properties of the boulders were kept the same and we employed the tensile strength and fracture model as described in \cite{Jutzi2008,Jutzi2015}, with parameters corresponding to a tensile strength $Y_T \approx$ 20 MPa for a $\approx$ 2 cm specimen. For the boulders on Dimorphos, the average $Y_T \approx$ 10 MPa (Table~\ref{table:parameters}). However, based on previous impact studies \cite[e.g.,][]{ormo2022}, we do not expect the impact outcome to be very sensitive to the boulder tensile strength.

The target matrix material response to shear deformation is described by a simple pressure-dependent strength model \cite{Lundborg1967, Collins2004}. The ability of a material to resist different types of stresses is an indicator of its strength. Granular materials, for instance, may exhibit considerable shear strength due to the presence of van der Waals forces and the particles' inability to separate or slide over each other due to interlocking mechanisms \cite{Sanchez2014, Scheeres2019, Ferrari2020}. Here we focus on the influence of the shear strength at zero pressure, commonly known as cohesion (or cohesive strength). Another important term in the strength model is the coefficient of internal friction. Though this parameter cannot be directly determined, it is possible to relate it to the angle of repose (See Methods) and bound the range of plausible values by making reasonable assumptions. The angle of repose of low cohesion materials has been measured to be $\theta$ = 22$^\circ$ ($f$ = 0.4) for glass beads \cite{Lajeunesse2005}, $\theta$ = 30$^\circ$ ($f$ = 0.55) for quartz sand \cite{Lube2004} and $\theta$ = 35–-45$^\circ$ ($f$ = 0.7--0.9) for lunar regolith \cite{Mitchell1972}. 

On Dimorphos, for values of cohesion lower than $Y_0$ $\approx$ 4 Pa, the impact occurs in the gravity-dominated regime where crater growth is halted by the asteroid's small gravity, rather than its cohesion \cite{Cheng2022}. Therefore, we first model impacts into cohesionless rubble-piles (i.e., $Y_0$ = 0 Pa, but with a coefficient of internal friction of $f$ = 0.55 \cite{Lube2004}, which is equivalent to $\theta$ $\approx$ 30$^\circ$). Given our other assumptions regarding material properties, we consider that these models of impacts into targets with no cohesion provide an upper limit on the possible momentum enhancement that can be achieved from the DART impact for a given asteroid mass.

%%%% How we calculate beta %%%%%
The momentum enhancement, $\beta$, was calculated using two distinct methods. For the first method, $\beta$ was calculated by summing the momentum over all the SPH particles with ejection velocities larger than $v_{esc}$. For a given impact, the magnitude of ejecta momentum
in the direction of the net ejecta momentum is given by 
\begin{equation}
	{p_{ej}} = |\sum m_e \vec{v_{ej}}|,
\end{equation}
where $m_e$ and $\vec{v_{ej}}$ are the mass and velocity vector of individual SPH particles, respectively. The ${p_{ej}}$ calculation takes the gravitational influence of Dimorphos into account, however, it does not account for the gravitational influence of Didymos. The second method, described in \cite{Syal2016}, tracks the velocity magnitude of the asteroid centre of mass post-impact by summing the momentum of all material that remains below the escape velocity after the reaccumulation of the ejecta. The absolute difference in $\beta$ resulting from the two calculation methods is used in the error calculation of our reported $\beta$ values.

\section*{Results}
\begin{figure*}[ht!]
\centering
\includegraphics[width=\linewidth]{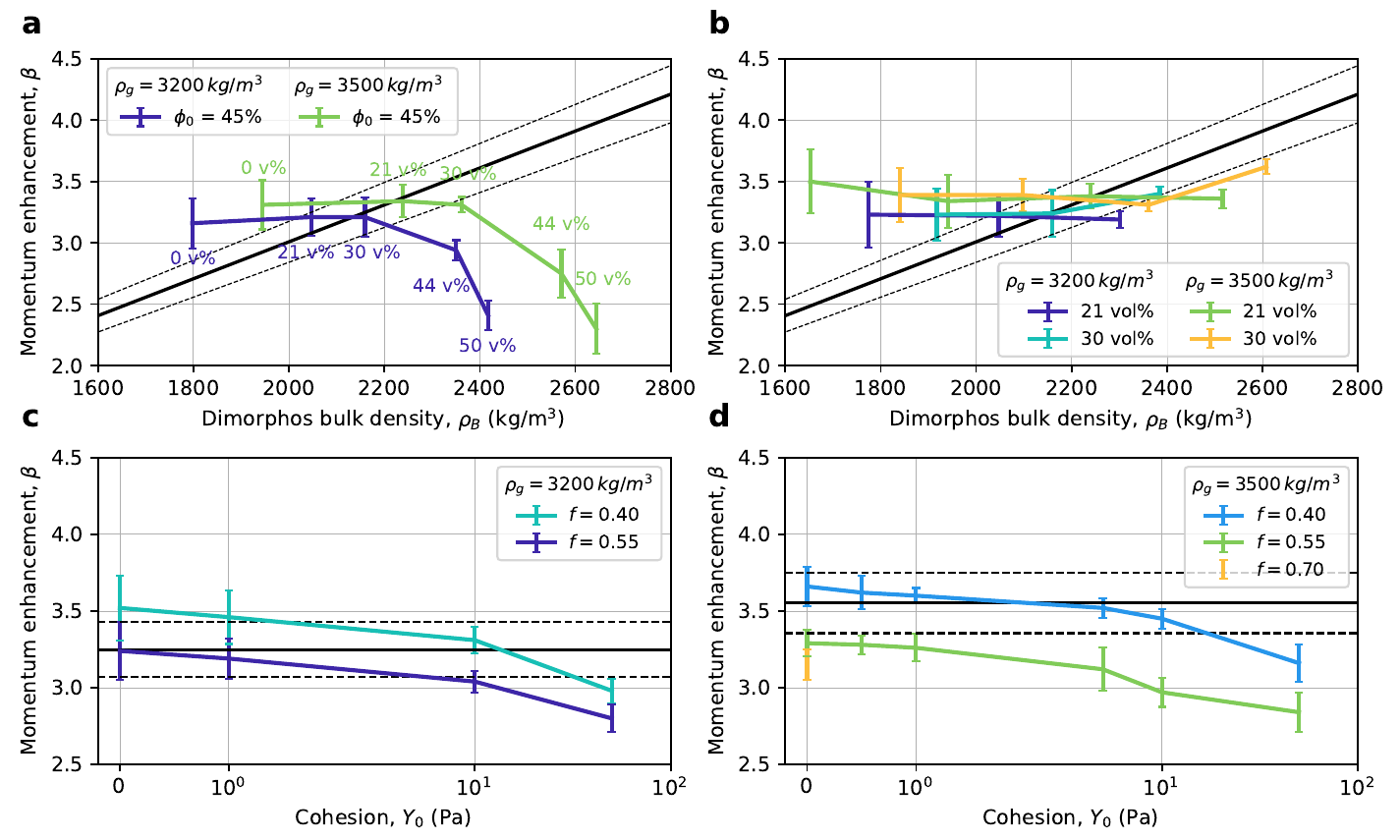}
\caption{Momentum enhancement, $\beta$, as derived from SPH simulations. The continuous black lines show the $\beta$ dependence on Dimorphos' bulk density, $\rho_B$, derived from dynamical simulations \cite{Cheng2023} (1-$\sigma$ uncertainty is shown by the dashed lines). Data points that cross the solid black line are consistent with the measured $\beta$ from the DART impact. The uncertainty on each simulation data (vertical bars) is calculated from the absolute difference in $\beta$ calculated from two different methods (see Momentum enhancement calculations; Methods). (a) $\beta$ as a function of $\rho_B$, for cohesionless targets with the same dimensions as Dimorphos \cite{Daly2023} and with boulder volume fractions ranging from 0 vol\% (no boulders larger than 2.5 m) to 50 vol\%. For fixed target volume (0.00181 km$^3$) and fixed matrix porosity ($\phi_0 = 45\%$), the mass and bulk density of Dimorphos vary with boulder packing. (b) $\beta$ as a function of $\rho_B$ for cohesionless targets with varying matrix porosity, between 35 and 65\% and two boulder packings: 21 and 30 vol\%. $\rho_B$ is calculated for a fixed asteroid volume and it varies with matrix porosity and boulder packing. (c) $\beta$ as a function of matrix cohesion ($Y_0$) for the DART impact into targets with varying coefficient of internal friction ($f$ = 0.4--0.55), an assumed grain density, $\rho_g$ = 3200 kg/m$^3$ and a 30 vol\% boulder packing. The horizontal line shows the $\beta$ derived from dynamical simulations for a target with $\rho_B$ = 2160 kg/m$^3$ (minimum density consistent with 1a and 1b results). (d) $\beta$ as a function of $Y_0$ for the DART impact into targets with $f$ = 0.4--0.70, $\rho_g$ = 3500 kg/m$^3$ and 30 vol\% boulder packing. The horizontal line shows the $\beta$ derived from dynamical simulations for a target with $\rho_B$ = 2360 kg/m$^3$ (maximum density consistent with 1a and 1b results).}
\label{fig:betas}
\end{figure*}

% Fig. 1a, boulder vol%
First, we vary the boulder volume fraction (i.e., the volume fraction of objects $>$2.5 m in size) within the target between 0 and 50 vol\%, while keeping the asteroid volume constant (Table~\ref{table:parameters}). As a result, the mass and the bulk density of the asteroid vary with boulder packing.
Our simulations show that $\beta$ is relatively insensitive to the boulder volume fractions up to $\approx$ 30 vol\% (Fig.~\ref{fig:betas}a). For boulder volume fractions larger than about 30 vol\%, the number of boulders much larger than the projectile close to the impact point is high enough that the crater efficiency and, subsequently, $\beta$ is drastically reduced by boulder interlocking \cite{Raducan2022d} and possibly armouring \cite{Tatsumi2018}, which hinders the crater growth. These results indicate that at least the surface and shallow subsurface of Dimorphos have a low volume fraction of boulders larger than 2.5 m (less than $\approx$ 40 vol\%), which is consistent with the last few images sent by DART before impact \cite{Daly2023}.

% Fig. 1b, porosity
Our simulation results for the DART impact on a cohesionless surface of Dimorphos also provide a means to constrain the bulk density of the asteroid using the measured $\beta$ (Fig.~\ref{fig:betas}b), assuming a grain density in the range of 3200--3500 kg/m$^3$. For a body with a fixed volume and grain density (Table~\ref{table:parameters}), the bulk density is influenced by its porosity. The dominant effect of additional target porosity is a reduction in bulk density of the target, rather than a reduction in ejecta from compaction of pore space. 
For example, decreasing the bulk density of Dimorphos will increase the overall crater size and decrease the asteroid's mass (for a fixed volume) and escape velocity. This allows for a greater total volume of ejecta to escape, but the ejecta mass is similar. Increasing the bulk density has the opposite effect. The consequence is that the total momentum of escaping ejecta measured constrains the target bulk density and porosity: for a cohesionless surface of Dimorphos, the upper bound on $\beta$ ($\approx$ 3.6) implies that the bulk density of Dimorphos is less than the current best estimate of the asteroid's bulk density of 2400 kg/m$^{3}$ \cite{Daly2023}. Thus, Dimorphos is likely more porous and therefore may have a rubble-pile structure throughout the whole body.

%%%%%%%%%%%%%%%%%%%%%%%%
% Cohesion, friction and beta
Since the surface strength of Dimorphos is poorly constrained, for a fixed boulder distribution (30 vol\%), matrix porosity ($\phi_0$ = 45\%) and grain density ($\rho_g$ = 3200 and 3500 kg/m$^{3}$) we vary the matrix material cohesion ($Y_0$ = 0--50 Pa) and coefficient of internal friction ($f$ = 0.4--0.7) (See Table~\ref{table:parameters}; Methods). 
Multiple possible combinations of cohesion, coefficient of internal friction, and bulk density could result in the observed deflection and account for the observed momentum enhancement (Fig.~\ref{fig:betas}c, d). Despite this non-uniqueness, it is possible to bound the range of plausible values by making reasonable assumptions since $f$ = 0.4 is a lower limit for geological materials. For a target with $f$ = 0.4 and $\phi_0$ = 45\% the cohesion on the surface of Dimorphos is likely lower than $\approx$ 50 Pa (Fig.~\ref{fig:betas}c, d). However, lower bulk densities ($\rho_B<2000$ kg/m$^3$) or higher matrix porosities ($\phi_0>55\%$) would require higher cohesion to match the observations (Fig.~\ref{fig:betas}).

%%%%%%%%%%%%%%%%%%%%%%%%%%%%%%%%%%%%%
%%%%%%%%%%%%%%%%%%%%%%%%%%%%%%%%%%%%%
\section*{Ejecta curtain opening angle and morphology}

\begin{figure*}[ht!]
\centering
\includegraphics[width=\linewidth]{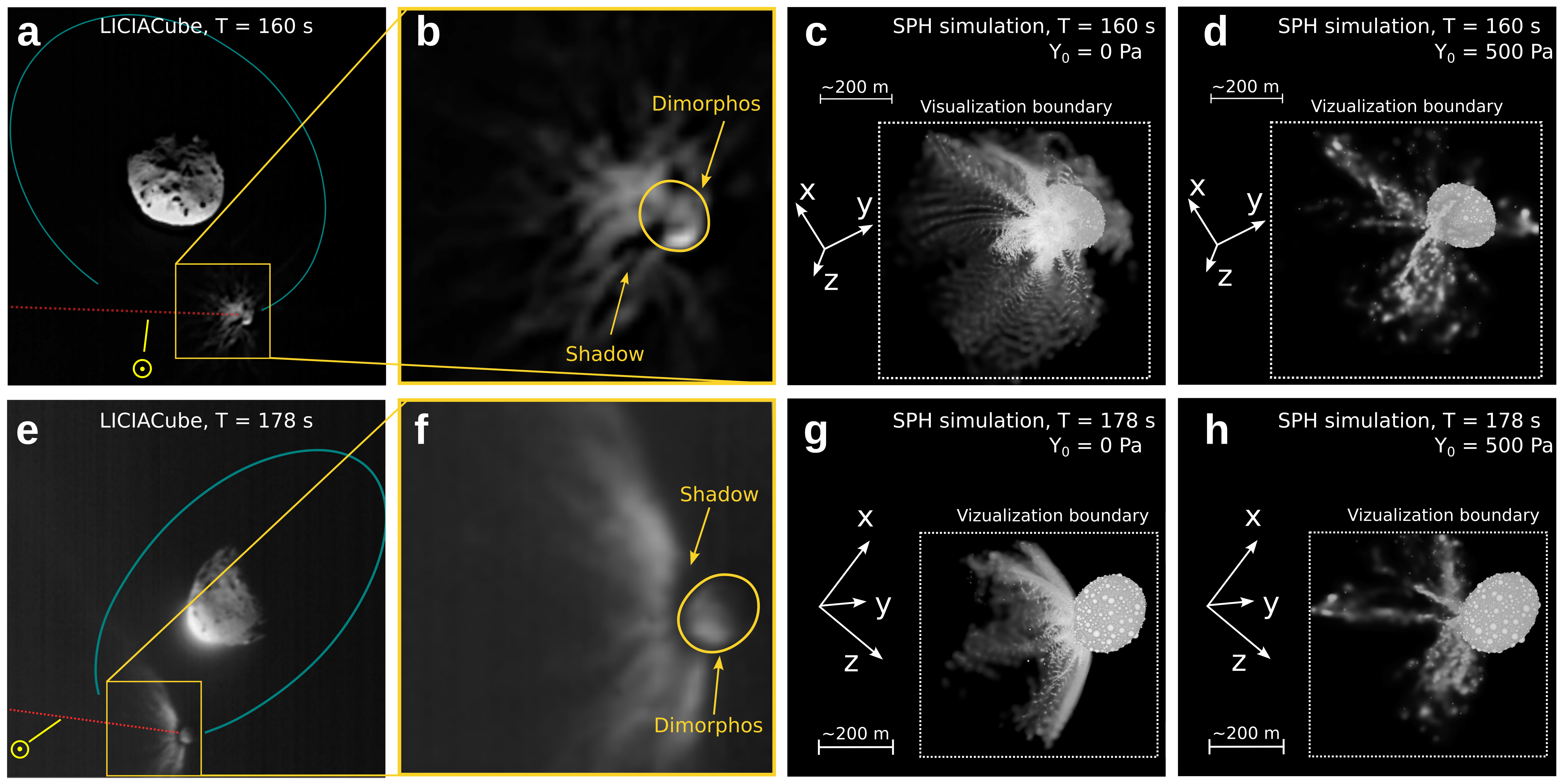}
\caption{LICIACube images of the expanding ejecta cone adapted from \cite{Dotto2023} compared with SPH simulation results. (a) Image acquired by the LUKE instrument onboard LICIACube at a distance of $\approx$ 76 km, 160 seconds after the impact. (b) Zoomed-in image of Dimorphos and impact ejecta. The approximate outline of the asteroid is shown in yellow. The ejecta curtain exhibits undulations, filamentary patterns, and shadows. (c) Bern SPH simulation of the DART impact into a cohesionless, rubble-pile Dimorphos-sized target (with $f$ = 0.55 and $\phi_0$ = 45\%), at $T$ = 160 s. Due to the limited visualisation domain, only a portion of the ejecta are shown. (d) Simulation of the impact into a rubble-pile Dimorphos-sized target with $Y_0$ =  500 Pa ($\beta$ = 2.26$\pm$0.28). (e) Image acquired by LUKE at a distance of $\approx$ 71 km, 178 seconds after the impact. (f) Zoomed-in image of Dimorphos and impact ejecta. (g) Same as (c) but at $T$ = 178 s. (h) Same as (d) but at $T$ = 178 s. The optical depth of the ejecta cone is not computed for this comparison between the observations and simulations output.}
\label{fig:LICIA}
\end{figure*}

\begin{figure*}[ht!]
\centering
\includegraphics[width=\linewidth]{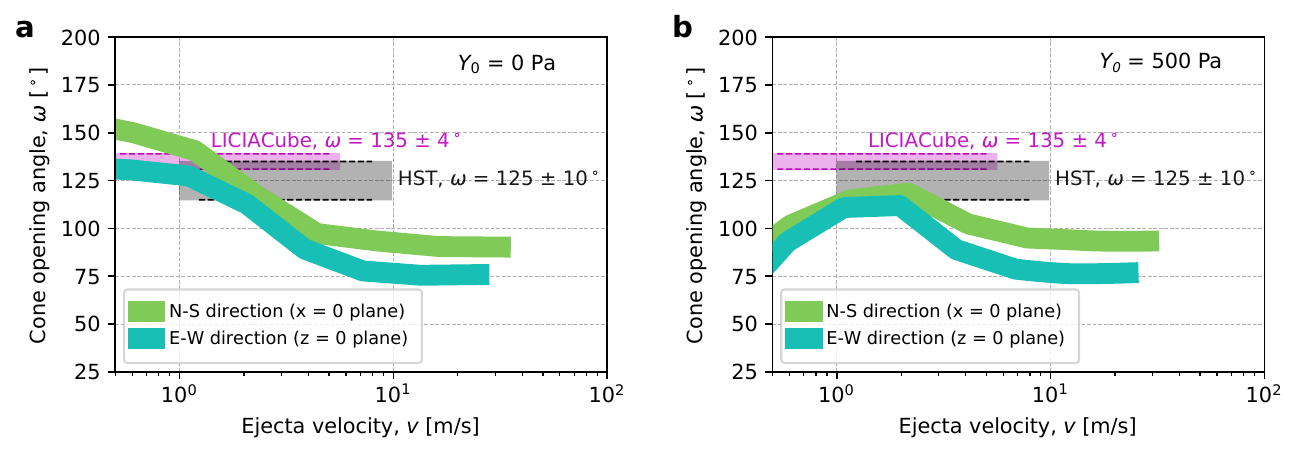}
\caption{Ejecta cone opening angle. (a) Cone opening angle derived from cohesionless ($Y_0$ = 0 Pa) SPH simulations (shown in Fig.~\ref{fig:LICIA}c and g) in the N-S direction ($x = 0$ plane) and in the E-W direction ($z$ = 0 plane). (b) Cone opening angle derived from SPH simulations with $Y_0$ = 500 Pa in the N-S direction ($x = 0$ plane) and in the E-W direction ($z$ = 0 plane). In both (a) and (b) the cone opening angle derived from observations is plotted for comparison: $\omega$ = 135 $\pm$ 4$^\circ$ is measured from LICIACube data based on the opening angle at the base of the cone at $T$ $\approx$ 170 s, resulting in ejecta velocities in the range of a few cm/s -- a few tens cm/s \cite{Dotto2023} and  $\omega$ = 125 $\pm$ 10$^\circ$ is measured in the Hubble Space Telescope (HST) data for ejecta in the range of 1--10 m/s \cite{Li2023}. Temporal evolution measurements of the observed ejecta cone is not possible due to the limited observation window.}
\label{fig:cone}
\end{figure*}

In our simulations of the DART impact into Dimorphos-like rubble pile targets, we observe the temporal changes of the ejecta cone opening angle and study dependences on target properties. We find no significant dependences of ejecta cone opening angle on the friction coefficient of the targets. This finding contrasts with the strong dependence of the cone opening angle on the coefficient of internal friction that is found in simulations \cite{Raducan2022c} of the DART impact into homogeneous planar targets.
Our present simulations of impacts into Dimorphos-like rubble piles find that the ejecta cone opening angle and ejecta mass depend on target cohesion. In Figures~\ref{fig:LICIA} and \ref{fig:cone} results for cohesionless targets ($Y_0$ = 0 Pa) are compared with those for cohesive targets ($Y_0$ = 500 Pa).

The ejecta plume for the cohesionless target (Fig.~\ref{fig:LICIA}c, g) is more massive than for the cohesive target (Fig.~\ref{fig:LICIA}d, h). For both cases, the fastest ejecta, released shortly after the impact with velocities higher than a few tens of m/s, form a cone opening angle $\omega \approx$ 90$^\circ$ (Fig.~\ref{fig:cone}a, b). Such fast ejecta are influenced by the spacecraft geometry (e.g., \cite{Raducan2022b, Owen2022}), which is highly simplified in this study. On the other hand, slower ejecta, released at late times after the impact (hundreds to thousands of seconds) with velocities $v_{esc} < v <$ 5 m/s, form a wider ejecta cone angle of $\approx$ 140$^\circ$ (at 1 m/s, Fig.~\ref{fig:cone}a) for the cohesionless target. For the cohesive target ($Y_0$ = 500 Pa), crater growth ceases about 100 seconds after the impact, before the crater grows large enough for the ejection angle to be influenced by target curvature. In this case, the maximum ejecta opening angle is $\approx$ 120$^\circ$ (at 1 m/s, Fig.~\ref{fig:cone}b). On the other hand, for the low cohesion target cases, the mass ejected at low velocities ($v_{esc} < v <$ 10 m/s) greatly exceeds the low velocity ejecta mass from the cohesive target case ($Y_0$ = 500 Pa), implying a larger cratering efficiency and crater growth continuing to later times, and resulting in a wider cone opening angle influenced by target curvature. 

The characteristics of the ejecta plume observed by LICIACube provide constraints regarding the target cohesion. At time after impact $T$ = 160 s, LICIACube’s LUKE acquired images that showed ejecta concentrated into rays which cast shadows along the ejecta cone (Fig.~\ref{fig:LICIA}a, b). At $T$ = 178 s, the bottom of the ejecta cone and the surface of Dimorphos are obscured by the shadow cast by the ejecta (Fig.~\ref{fig:LICIA}e,f). The shadow observed at $T$ = 178 s implies that crater growth and release of low speed ejecta continued to that time, consistent only with low cohesion target cases \cite[e.g.,][]{Holsapple2007}.

Images obtained from LICIACube \cite{Dotto2023} and Hubble Space Telescope observations \cite{Li2023} revealed a wide ejecta cone angle, estimated to be $\omega \approx$ 115–139$^\circ$. These observations determined the ejecta opening angle at specific times, up to 3 minutes after the impact for LICIACube \cite{Dotto2023} and up to 8 hr after the impact for Hubble \cite{Li2023}. To compare simulations results with observations of the ejecta opening angle at a specific time after impact, we determine the implied ejecta velocity at the base of the visible cone in the LICIACube images (\cite{Dotto2023}; Fig.~\ref{fig:LICIA}e, f) using its distance above the surface and time after impact (Fig.~\ref{fig:cone}).
	 
Overall, we find that to qualitatively reproduce the amount of material observed in the ejecta cone (Fig.~\ref{fig:LICIA}c, g), as well as the observed cone opening angle of up to $\omega \approx$ 139$^\circ$, Dimorphos’ surface cohesion must not exceed $\approx$ 500 Pa. From our suite of numerical simulations with the assumed boulder packing, matrix porosity, and grain density summarised in Table~\ref{table:parameters}, we find the target case with $f$ = 0.55, $\rho_B$ = 2200 kg/m$^{3}$, and $Y_0$ less than a few Pa is consistent with target mechanical properties inferred from surface geology \cite{Ernst2023} and produces a $\beta$-value (Fig.~\ref{fig:betas}), excavation timescale (Fig.~\ref{fig:LICIA}) and ejecta opening angle (Fig.~\ref{fig:cone}) most consistent with observations.

\section*{Deformation}

\begin{figure*}[ht!]
\centering
\includegraphics[width=\linewidth]{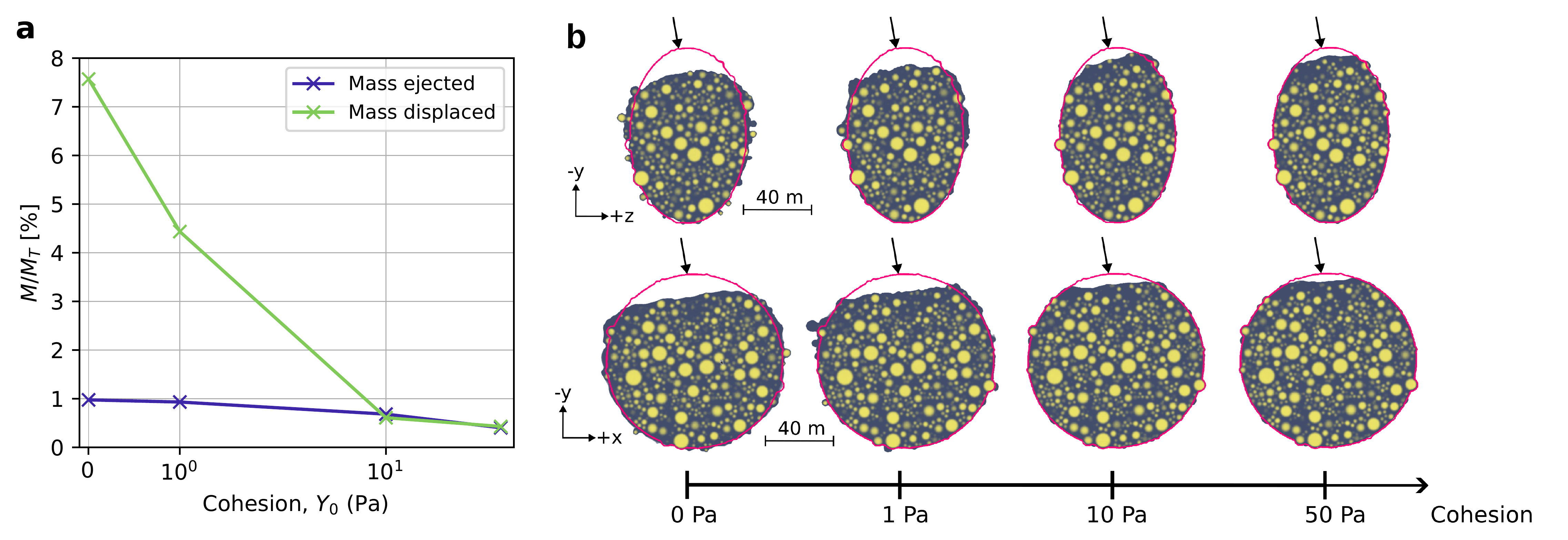}
\caption{Global deformation of Dimorphos. (a) Total target mass ejected with speeds above $v_{esc}$ and total target mass displaced or ejected below $v_{esc}$, normalised by the initial target mass, $M_T$, for a Dimorphos-like target with $f$ = 0.55, $\phi_0$ = 45\% and cohesion levels between $Y_0$ 0 and 50 Pa. (b) Two-dimensional slices (taken at $x$ = 0 in the $y-z$ plane and at $z$ = 0 in the $x-y$ plane), at $T \approx$ 1 hour after the impact. The boulder material is shown in yellow and the matrix material is shown in blue. The red contour shows the initial target profile, before the impact. The black arrows show the impact direction.}
\label{fig:deformation}
\end{figure*}

Observations from the first few hours after impact imply that more than 1.3--2.2 $\times$ 10$^7$ kg of ejecta were released from the DART impact (equivalent to 0.3--0.5\% of Dimorphos's mass, assuming a bulk density of 2400 kg/m$^3$) \cite{Graykowski2023}. Our simulation results for weak ($Y_0 <$ 50 Pa), Dimorphos-like targets ($f = 0.55$ in Fig.~\ref{fig:betas}c) show that the amount of ejected material could be as high as 1\% of Dimorphos' mass (Fig.~\ref{fig:deformation}a). At the same time, up to 8\% of Dimorphos' mass may have been displaced or ejected below the escape velocity of Dimorphos. In all impact scenarios simulated here, the DART impact does not produce a conventional impact crater and instead causes global deformation of the target (Fig.\ref{fig:deformation}b). 

The outcome of the impact in terms of the post-impact target morphology is highly sensitive to the target cohesion. For a cohesionless target, the ratio of the major to intermediate axes, $a/b$, could have changed from the reported pre-impact value of 1.02 $\pm$ 0.02 \cite{Daly2023} to as much as 1.2. Such a large change in the $a/b$ ratio is detectable with the highest-quality post-impact lightcurve data \cite{Pravec2016, Pravec2022}.  

% Implications for Dynamics
Global deformation of Dimorphos modifies the gravitational field between Didymos and Dimorphos and leads to significant implications for its orbit. The shape change would cause an additional perturbation to Dimorphos’s orbit, on top of those caused by the spacecraft momentum and ejecta recoil, and this effect can account for a few seconds to several minutes of the observed orbit period change (i.e., $\sim$ 33 minutes), depending on the magnitude of the deformation \cite{Nakano2022}. Any deformation would alter Dimorphos's mass distribution and affect its post-impact rotation state (e.g., \cite{Agrusa2021, Richardson2022}). 

\section*{Implications for binary asteroid system formation} 

Our numerical simulations suggest that Dimorphos is likely a rubble-pile asteroid with a bulk density comparable to or lower than that of Didymos. Our calculations based on the observed boulder SFD indicate that the macroporosity estimate for the surface of Dimorphos ($\approx$ 35\%) is approximately twice the value obtained for the surface of Ryugu, as determined through the same method \cite{Grott2020, Tricarico2021}, but comparable with the macroporosity on Itokawa \cite{Fujiwara2006, Grott2020}. However, it is worth noting that this estimate is only a rough approximation due to the limited data currently available and the upcoming Hera mission will be able to provide better constrains. 

Our findings serve as crucial evidence regarding the origin of Dimorphos as a secondary in a double asteroid system. To maintain its structural stability given its rapid rotation period of 2.26 hours, the primary, Didymos, likely requires a higher cohesive strength, estimated to be on the order of 10s Pa \cite{Zhang2021}. This level of cohesion can be attributed to van der Waals forces acting between the fine regolith grains \cite{Scheeres2010}, or to a coherent inner core \cite{Ferrari2022}. However, our best-fit scenarios indicate that Dimorphos, the satellite of Didymos, exhibits a cohesive strength of less than a few Pa. This observed disparity in cohesive strength between Didymos and Dimorphos suggests a potential scarcity of fine grains within Dimorphos' structure as well as a weak and fragmented internal structure.

The material properties estimated in our study support the hypothesis that Dimorphos formed through rotationally or impact-induced mass shedding and subsequent re-accumulation from Didymos. The accretion of orbiting mass shed from Didymos occurs over a period of several days to years \cite{Walsh2008}, during which fine grains tend to escape from the system due to solar radiation pressure \cite{FerrariRaducan2022}. As a result, the accreted satellite, Dimorphos, has limited fines and cohesion. 

While the mechanical properties of Dimorphos resemble those of Ryugu and Bennu (e.g., \cite{Arakawa2020, Jutzi2022, Walsh2022, Barnouin2022}), these findings may not be applicable to single small S-type asteroids, specifically to their moons. The implications of our study may extend beyond Dimorphos and provide valuable insights into the formation processes of similar small S-type binary asteroid systems. 

Moreover, since the DART spacecraft likely caused global deformation of Dimorphos, we can infer that similarly formed asteroid moons are easily reshaped and their surfaces are relatively young \cite{Raducan2022}. Overall, the findings of this study provide valuable information for understanding the formation and characteristics of binary asteroids, and will inform future exploration and asteroid deflection efforts.

%%%%%%%%%%%%%%%%%%%%%%%%%%%%%%%%%%%%%
%%%%%%%%%%%%%%%%%%%%%%%%%%%%%%%%%%%%%
%%%%%%%%%%%%%%%%%%%%%%%%%%%%%%%%%%%%%

\newpage

\section*{Data Availability}
Additional supporting information and input data for the model simulations used in this work is archived on GitHub (doi:10.5281/zenodo.10246671).

\section*{Code availability}
A compiled version of the Bern SPH code, as well as the necessary input files are available from the corresponding author upon request. SPH data visualisation was produced using the NCAR Visualization and Analysis Platform for Ocean, Atmosphere, and Solar Researchers (VAPOR version 3.8.0) [Software] (doi:10.5281/zenodo.7779648).

\section*{Acknowledgments}
We thank Jessica Sunshine, Mallory DeCoster, Dawn Graninger, Jason Pearl, Angela Stickle and the rest of the DART Impact Working Group for the constructive discussions. 

SDR and MJ acknowledge support by the Swiss National Science Foundation (project number 200021\_207359). This work was supported by the DART mission, NASA Contract No. 80MSFC20D0004. GSC and TMD acknowledge support from UK Science and Technology Facilities Council Grant ST/S000615/1. FF acknowledges funding from the Swiss National Science Foundation (SNSF) Ambizione grant No. 193346. KMK, MBS, JMO Portions of this work were performed by Lawrence Livermore National Laboratory under DOE Contract DE-AC52-07NA27344. LLNL-JRNL-846795, PM acknowledges financial support from the CNRS through the MITI interdisciplinary programs through its exploratory research program an, from ESA and from the University of Tokyo. PM, RL, KW, NM and CQR acknowledge the support from the European Union’s Horizon 2020 research and innovation program, grant agreement No. 870377 (project NEO-MAPP). NM, CQR and PM	acknowledge funding support from the Centre National d’Etudes Spatiales (CNES). RN acknowledges support from NASA/FINESST (NNH20ZDA001N/80NSSC21K1527). ED, EME, PHH, SI, AL, MP, AR and FT	acknowledge financial support from Agenzia Spaziale Italiana (ASI, contract No. 2019-31-HH.0). MP, AL and FT also acknowledge support from ASI, contract No. 2022-8-HH.0. Work by EGF was carried out at the Jet Propulsion Laboratory, California Institute of Technology, under a contract with the National Aeronautics and Space Administration ($\#$80NM0018D0004). JO acknowledge support by grant PID2021-125883NB-C22 by the Spanish Ministry of Science and Innovation/State Agency of Research MCIN/AEI/ 10.13039/501100011033 and by ‘ERDF A way of making Europe’. JO, IH, SR, MJ, RL, KW	acknowledge support by Consejo Superior de Investigaciones Científicas (CSIC) (project ILINK22061). The work by PP was supported by the Grant Agency of the Czech Republic, grant 20-04431S.

\section*{Author contributions}

SDR, MJ, AFC conceptualised the study. 
SDR ran the simulations and analysed the data.
SDR, MJ, AFC, OB, GSC wrote the initial draft. 
YZ provided rubble-pile models.
RTD, CME, OB provided the shape model of Dimorphos. 
TLF provided the viewing geometry for comparison to LICIACube data.
MH, JYL, PHH provided measurements of the ejecta.
YZ, FF, HFA helped with the interpretation of results. 
RN helped with the deformation affects on dynamics. 
MP, AL, FT provided boulder SFDs.
CQR, NM provided the boulder shapes.
AFC, HFA, BWB provided momentum enhancement measurements.
AR, ED, PHH provided LICIACube measurements. 
PP provided observational inputs.
PS helped with the interpretation of cohesion. 
TMD, KMK, PM, MBS, NLC, ED, EGF, IH, SI, RL, JO, MO, ASR, KW, AZ, EME provided comments that substantively revised the manuscript.
CM, BHM provided the stereographs of the Didymos system.

\section*{Competing interests}
The authors declare no competing interests

\section*{Tables}

\begin{table}[h]
 \centering 
	%\small
	\caption{Table of fixed and varied target parameters.}
	\vspace{0.3cm}
\begin{tabular}{ll}
Fixed parameters & \\
\hline
Target dimensions \cite{Daly2023} & 177 $\times$ 174 $\times$ 116 m\\
Target volume \cite{Daly2023} & 0.00181 km$^3$ \\
Boulders SFD \cite{Pajola2023} & see Methods \\
Boulders at impact location \cite{Daly2023} & see Methods \\
Boulder tensile strength & 10 MPa \\
Boulder porosity &		10\%  \\
 &		  \\
Varied parameters & \\
\hline
Boulder packing & 0 -- 50 vol\%\\
Grain density, $\rho_g$   & 3200/3500 kg/m$^3$\\
Matrix porosity, $\phi_0$ & 35 -- 65\%\\
Matrix cohesion, $Y_0$ & 0 -- 500 Pa\\
Matrix internal friction coeff., $f$ & 0.4 -- 0.7 
\end{tabular}
\label{table:parameters}
\end{table}

\section*{Figure Legends/Captions}

%%%%%%%%%%%%%%%%%%%%%%%%%%%%%%%
%%%%%%%%%%%%%%%%%%%%%5

%%%%%%%%%%%%%%%%%%%%%%%%

%%%%%%%%%%%%%%%%%%%%%%%%%%%%%%%%%%%%%
%%%%%%%%%%%%%%%%%%%%%%%%%%%%%%%%%%%%%

%%%%%%%%%%%%%%%%%%%%%%%%%%%%%%%%%%%%%

%\bibliographystyle{naturemag}
%\bibliography{bibdata.bib}
\printbibliography

\newpage

\clearpage

\end{document}